\documentstyle[aps,twocolumn, floats]{revtex}

\def\lb{\label}
\def\nF{S_{\parallel}}

\begin{document}

\draft

\twocolumn[\hsize\textwidth\columnwidth\hsize\csname@twocolumnfalse\endcsname

\title{Guiding Neutral Atoms with a Wire}

\author{Johannes Denschlag \and Donatella Cassettari \and
J\"org Schmiedmayer}

\address{Institut f\"{u}r Experimentalphysik, Universit\"{a}t Innsbruck, A-6020 Innsbruck,  AUSTRIA}

\date{\today}
\maketitle
\begin{abstract}
We demonstrate guiding of cold neutral atoms along a current
carrying wire. Atoms either move in Kepler-like orbits {\em around}
the wire or are guided in a potential tube {\em on the side} of the
wire which is created by applying an additional homogeneous bias
field. These atom guides are very versatile and promising for
applications in atom optics.

\end{abstract}

]


\narrowtext

In atom optics \cite{AtomOptics} it is usually desirable to
separate atoms as far as possible from material objects in order to
obtain pure and isolated quantum systems. With cooling and trapping
techniques \cite{LasCool} being well established, there is now an
interest in bringing the atoms close to material macroscopic
objects. The proximity of the atoms to the object allows the design
of tailored and easily controllable potentials which can be used to
build novel atom optical elements.

In this letter we demonstrate two simple and versatile atom guides
that are based on magnetic trapping potentials created by a thin
current carrying wire:  The `Kepler guide' and the `side guide'. In
our experiments we study the transport of cold lithium atoms from a
magnetic-optical trap in these guiding potentials. We were able to
measure scaling properties and extract characteristic atomic
velocity distributions for each guide. The `side guide' is
especially interesting because it can easily be miniaturized and
combined with other guides to form mesoscopic atom optical
networks.

We start with  discussing the interaction of a neutral atom and a
current carrying wire and then describe our guiding experiments.

{\bf Kepler guide: }The magnetic field of a rectilinear current $I$
is given by:
\begin{equation}
        B = \frac{\mu_{0}}{2 \pi } \frac{I}{r}  \hat{e}_{\varphi} ,
        \label{MagField}
\end{equation}
where $\hat{e}_{\varphi}$ is the circular unit vector in cylindrical
coordinates.  An atom with total spin $\vec{\bf S}$ and magnetic
moment $\vec{\mu} = g_S \mu_B \vec{\bf S}$ experiences the
interaction potential $V_{mag}= -\vec{\mu} \cdot \vec{\bf B} = - g_S
\mu_B \nF B$, where $\nF$ is the projection of $\vec{\bf S}$ on $\vec{\bf
B}$. In general the vector coupling $\vec{\mu} \cdot \vec{\bf B}$
results in a very complicated  motion for the atom. However, in our
experiments  the Larmor precession ($\omega_{L}$) of the magnetic
moment is much faster than the apparent change of direction of the
magnetic field in the rest frame of the atom ($\omega_{B}$) and an
adiabatic approximation can be applied. $\nF$ is then constant and
the atom can be described as moving in a scalar 1/r potential.
\begin{figure}[tbh]
    \begin{center}\hspace{-5mm}\mbox{\input epsf \epsfxsize\columnwidth\epsfbox{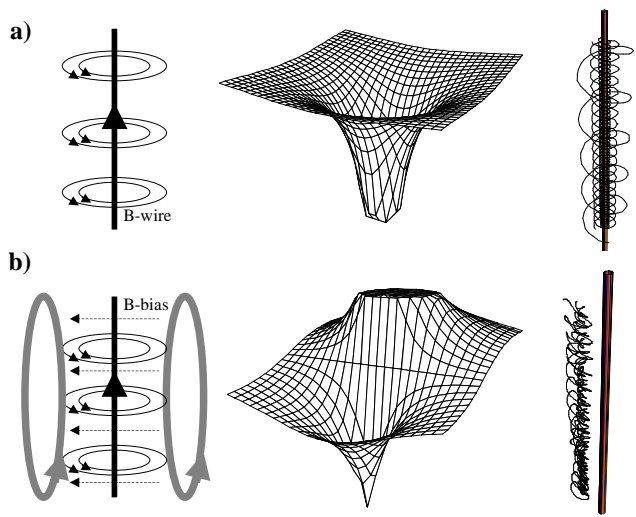}}\end{center}
    \caption{Two configurations for guiding  neutral atoms
        with a current carrying wire.  The left graphs display the
        respective magnetic field configurations.
        The middle shows the corresponding guiding potentials and on
        the right hand side typical calculated trajectories of guided atoms are drawn.
        {\bf a)} Guiding atoms in their {\em high field seeking} state, where
        atoms circle in Kepler orbits {\em around} the wire.
        {\bf b)} Guiding atoms in their {\em low field seeking} state along a line of
        a magnetic field minimum.
        Atoms are guided {\em on the side} of the wire.}
        \label{f:cw_guides}
\end{figure}
For $\vec{\bf \mu}$ ``parallel'' to $\vec{\bf B}$, ($\vec{\mu} \cdot
\vec{B} > 0$), the atom is in its {\em high field seeking state}, and
the interaction between the atom and the wire is attractive (see
Fig.~\ref{f:cw_guides}). The atoms in this state can  be trapped
and move in Kepler-like orbits around the wire
\cite{Vlad61,Schm92,Schm96a,CuWireTheory}.

{\bf Side guide: }Combining the field of the current carrying wire
with a homogeneous magnetic  bias field $B_b$ perpendicular to the
wire breaks the rotational symmetry resulting in unstable orbits
for the strong field seeking atoms \cite{BiasWire}. In addition the
bias field has the effect of exactly canceling the circular
magnetic field of the wire along a line parallel to the wire at a
distance $r_s = (\mu_0/2\pi)(I/B_b) $. Around this line the
magnetic field increases in all directions and forms a tube with a
magnetic field minimum in its center. Atoms in the {\em low field
seeking state} ($\vec{\mu}\cdot\vec{B} < 0$) can be trapped in this
tube and guided along the wire as shown in Fig.~\ref{f:cw_guides}b.

\begin{figure}
    \begin{center}\hspace{-5mm}\mbox{\input epsf \epsfxsize\columnwidth\epsfbox{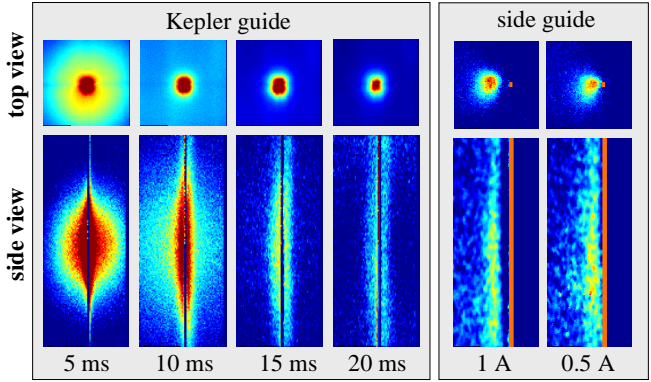}}\end{center}
    \caption{(color)
     Guiding of atoms along a current carrying wire.
     Pictures of the atomic clouds taken from above and from the side are shown.
     {\em (left)}  Series of pictures illustrating the guiding of atoms
     along the current carrying wire  in their
     {\em strong field seeking} state (Kepler guide).
     For times shorter than 15 ms
     the expanding cloud of untrapped atoms is also visible.
     {\em (right)} By adding a bias field atoms in the {\em weak field seeking}
     state are guided {\em on the side} of the wire (20 ms guiding time).
     Choosing different currents through the wire (here 0.5 A and 1 A)
     the distance of the side trap from the wire can be controlled.
     The location of the wire is indicated with a orange line (dot).
     The pictures show a 2 cm long section of the wire that is illuminated
     by the trapping beams.}
    \lb{f:WireGuideExp}
\end{figure}

Our experiments investigating
the Kepler- and side guides are carried out in four steps:

{\em (a)} First we load about $2\times10^7$ lithium atoms into a
magnetic optic trap  (MOT) \cite{LiMOT}, displaced typically 1 mm
from a $50
\;
\mu$m thick and $10 \; cm$ long tungsten wire. The MOT is loaded at
a distance to the wire in order to prevent trap losses due to atoms
hitting the wire \cite{Den98}.

{\em (b)} After loading the trap we shut off the slower beam and
shift the atoms within $5$ ms to the position where they are loaded
into the atom guide. This shifting is done by applying an
additional magnetic offset field and moving the center of the
magnetic quadrupole field, which defines the position of the MOT.
Simultaneously the frequency and intensity of the trapping lasers
are changed to control the size  and temperature   of the atom
cloud (typically 1.6 mm diameter(FWHM) and T $\sim$ 200 $\mu$K
which corresponds to a velocity of about 0.5 m/s).

 {\em (c)}
We then release the atoms from the MOT  by switching off the laser
light, the MOT magnetic fields, and the shifting fields within $
0.5$ ms. At this point the current through the wire (typically 1 A)
and, if desired, a bias magnetic field is switched on within $
100\;\mu $s. From then on the atoms move in the tailored guiding
potential. Starting from an initially well localized  atom cloud
the density distribution expands and changes shape according to the
forces on the atoms.

{\em (d)} After a given guiding/trapping time, the spatial
distribution of the atoms is measured by imaging the fluorescence
from optical molasses \cite{LasCool} using a CCD camera. For this
the guiding fields (current through the wire and bias field) are
switched off and molasses laser beams are switched on for a short
time (typically $< 1 \; $ms). Pictures are taken from above
(looking in wire direction) and from the side (looking onto the
wire from an orthogonal direction). This allows to study both, the
radial confinement and the guiding of the atoms along the wire.

Typical pictures of atoms orbiting around the wire (Kepler guide)
and being guided on the side of the current  carrying wire (side
guide) are shown  in Fig.~\ref{f:WireGuideExp}. The left set of
graphs visually demonstrates loading and guiding of atoms with the
Kepler guide: The atoms are released from the MOT at $t=0$  in the
center of the wire. Some fraction of these atoms will be bound by
the guiding potential, the rest  forms an expanding cloud that
quickly fades away within about 15 ms. The bound atoms are guided
along the wire corresponding to their initial velocity component in
this direction.  Consequently a cylindrical atomic cloud forms
around the wire that expands along the wire. For long guiding times
the bound atoms leave the field of view, and the fluorescence
signal of the atoms decreases. The top view images show a round
atom cloud that is centered on the wire suggesting that atoms
circle around the wire.

The graphs on the right hand side of Fig.~\ref{f:WireGuideExp} show
atoms that are bound to the side guide after 20 ms of guiding time.
In the given examples two different currents (1 A and 0.5 A) were
sent through the wire. The distance of the guide from the wire
changes clearly with the current.

\begin{figure}
    \begin{center}\hspace{-5mm}\mbox{\input epsf \epsfxsize\columnwidth\epsfbox{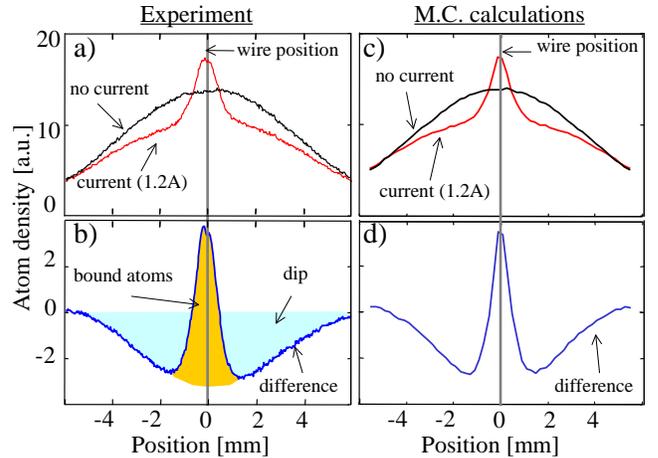}}\end{center}
    \caption{(color)
    Radial atomic density distribution for the Kepler guide after
    7.5 ms of interaction time. {\em (left)} measurement,
    {\em (right)} calculation. The upper graphs show the atomic distributions
     for a current of 1.2 A  through the wire and without current (free expansion).
    The difference between the two distributions is shown below.
    The central peak is due to
    2 dimensional trapping of atoms in the guide. The orange shaded
    area, determined by fitting the sum of two Gaussians to the data,
    is a measure of the number of trapped atoms.}
    \lb{f:MTech}
\end{figure}


The CCD pictures can be used for further analysis as illustrated in
figure~\ref{f:MTech} for the Kepler guide.
For this the CCD images are integrated yielding a projection of the
density distribution of the atoms in a direction perpendicular to
the wire. Figure~\ref{f:MTech}a shows two such distributions: One
corresponding to free expansion of the atomic cloud with no current
through the wire. This yields a Gaussian distribution which is
typical for atoms released from a MOT. The other one, corresponding
to atoms interacting with a current carrying wire, exhibits a
pronounced peak centered around the wire which can be attributed to
trapped atoms orbiting the wire. The peak sits on top of the broad
distribution of non-trapped atoms. In order to extract only the
effects of the magnetic guiding potential on the atomic cloud, the
two curves are subtracted from each other (Fig.~\ref{f:MTech}b).
The peak of the trapped atoms now sits in a broader dip, which is
caused by repulsion of atoms in low field seeking states from the
wire and by the fact that atoms trapped around the wire are now
missing from the expanding atomic cloud. Our experimental data
agree well with numerical simulations of the atomic density
distributions as shown in Fig. \ref{f:MTech}c,d.

\begin{figure}
    \begin{center}\hspace{-5mm}\mbox{\input epsf \epsfxsize0.9\columnwidth\epsfbox{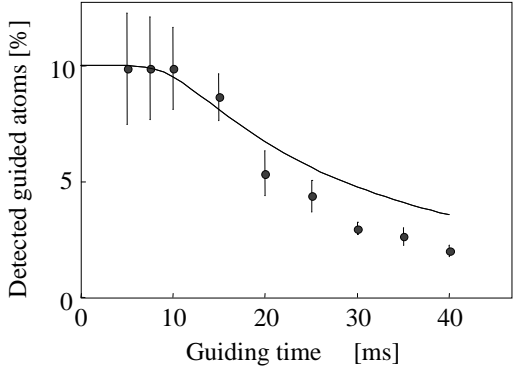}}\end{center}
    \caption{ Detected
    fraction of the MOT-atoms that are bound to the Kepler
 guide:  Atoms are only detected as long as they are located in the field of view of the
 CCD camera.
 The solid line
results from a calculation where we take into account the finite
size of the laser beam that is used to illuminate the atoms. The
additional loss at higher interaction times is most likely due to
uncompensated bias fields. }
    \lb{f:fig4}
\end{figure}

From fits to these experimental data we can extract quantitative
information on the guiding of atoms. As an example
Fig.~\ref{f:fig4} shows the number of detected atoms that are bound
to the Kepler guide as a function of guiding time. The data is
given as the fraction of the total number of atoms of the MOT. The
observed number of guided atoms decreases with time mainly because
the expanding atomic cloud leaves the detection region given by the
2 cm diameter laser beams. The solid line represents a
corresponding calculation of how the atomic signal is expected to
fall due to the free expansion of the atomic cloud along the wire
and the falling under the influence of gravity. We attribute the
additional loss observed in the experiment to magnetic stray fields
that render the atomic orbits unstable, i.e. atoms will hit the
wire and are lost. There will also be a small contribution to the
losses because of a decrease in the wire current over time, caused
by the increasing resistance due to ohmic heating of the wire.

From Fig.~\ref{f:fig4} we can also  extract the absolute loading
efficiency for atoms from the MOT into the guide. In  agreement
with Monte Carlo calculations we find loading efficiencies in the
range of 10 \% for the {\em Kepler guide}.

Using the same wire current, the loading of the {\em side guide} is
less efficient (up to $4\%$ loading efficiency), mainly due to the
smaller depth and `size' of the side guide potential.

\begin{figure}
    \begin{center}\hspace{-5mm}\mbox{\input epsf \epsfxsize0.85\columnwidth\epsfbox{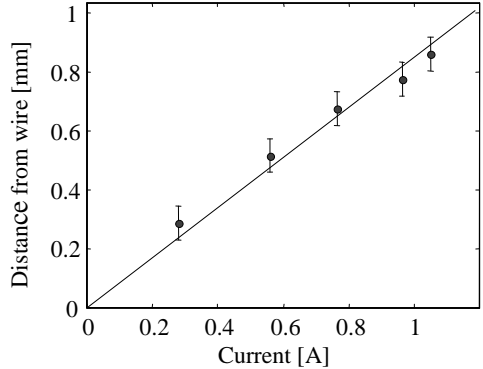}}\end{center}
    \caption{
    Position of guided atoms vs. the current through the wire for
    the side guide.  Decreasing the current brings the trap
    closer to the wire and consequently makes it smaller and steeper.  }
    \lb{f:Compress}
\end{figure}

However, the side guide exhibits  interesting scaling properties:
Its trap depth is given by the {\em magnitude} of the bias field.
With a fixed  trap depth the trap size and its distance from the
wire can be controlled by the current in the wire. The paradoxical
situation arises that the trap gets smaller (size $\propto I/B_b$)
and steeper (gradient $\propto B_b^2 / I$) for {\em decreasing}
current in the wire. The smallest and steepest trap achievable with
a fixed bias field is only limited by the requirement that it must
be located outside the wire. A simple calculation shows for example
that a trap with a gradient of over 1000 Gauss/cm can be achieved
with a moderate current of 0.5 A and an offset field of 10 Gauss.
The trap would be located 100 $\mu$m away from the wire center.
Figure~\ref{f:Compress} illustrates the linear scaling of $r_s$,
the distance of the side guide from the wire,  as a function of the
wire current (see also the right hand side of Fig.
\ref{f:WireGuideExp}).

Other interesting information about the transverse confinement can
be extracted by measuring the momentum distribution of the trapped
atoms.  This can be accomplished by ballistic expansion after
switching off the guiding potentials. After a few ms of expansion
the spatial distribution well represents the velocity distribution
of the atoms. Figure~\ref{f:fig6} shows the spatial atomic
distribution 9 (7) ms after switching off the guides. A clear
distinction can be seen between the two types of guides. Atoms in
the Kepler guide, where atoms circle around the wire, expand in a
ring (Fig.~\ref{f:fig6}a), showing clearly that there are no
zero-velocity atoms. This is because in order to be trapped in
stable orbits {\em around} the wire the atoms need sufficient
angular momentum and therefore velocity. Atoms with too little
angular momentum hit the wire and are lost. The low field seeker
atoms of the side guide, however, do not have this constraint.
Their velocity distribution is a standard Gaussian
(Fig.~\ref{f:fig6}b).

\begin{figure}
    \begin{center}\hspace{-5mm}\mbox{\input epsf \epsfxsize\columnwidth\epsfbox{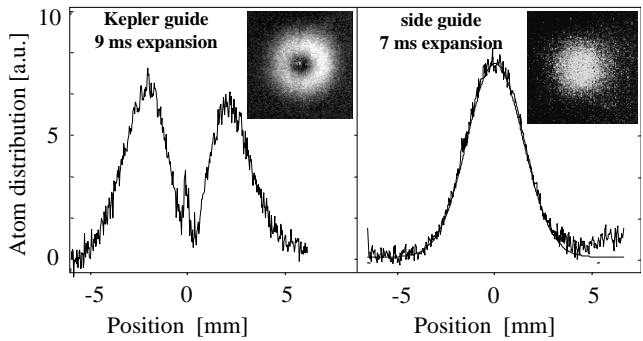}}\end{center}
    \caption{
    Atomic distribution after free expansion of 9(7) ms for atoms
    that have been guided along the wire. The two
    profiles that are shown are cuts through
    the center of the respective CCD images
    in the right upper corner.
    For the high field seeker trap {\em (left)} the expanded cloud is doughnut-shaped
    due to the orbital motion of the atoms around the wire. For
    guiding on the side {\em (right)} we obtain a standard Gaussian distribution.
    }
    \lb{f:fig6}
\end{figure}

In conclusion we presented two novel methods to build guides for
cold atoms, by using potentials created by a current carrying wire:
Guiding {\em strong field seekers} in Kepler-like orbits around a
wire and guiding {\em weak field seekers} in a potential tube along
the side of a wire. These atom guides  represent promising
techniques for future applications in atom optics, because of their
simplicity and versatility. For example by combining different
wires we can   construct beam splitters \cite{test},
interferometers and more complex matter wave networks
\cite{Matter_Guides}. Achieving the ultra high vacuum (UHV)
conditions required for coherent guiding is very simple, since the
propagation of atoms is in the open and {\em not} in an enclosed
space like for hollow optical fibers \cite{HollowFibers}.

In addition side guides can  be mounted  on a surface, so that
atoms are guided above the surface along the wires. This renders
the wires more stable and at the same time  allows for efficient
cooling which enables also thin wires to support sizeable currents
(see also \cite{SurfaceMounting}). Therefore we believe to have now
a method for miniaturization and integration of many atom optical
elements into one single quantum circuit in the near future,
creating mesoscopic atom optics similar to mesoscopic quantum
electronics.

We thank A. Zeilinger for his generous support throughout the work.
This work was supported by the Austrian Science Foundation (FWF),
project S065-05, the Jubil\"{a}ums Fonds der \"{O}sterreichischen
Nationalbank, project 6400, and by the European Union, contract Nr.
TMRX-CT96-0002.


\begin{thebibliography}{99}
\bibitem{BiasWire}
With the rotational symmetry broken, angular momentum is no longer
conserved. This results in unstable orbits for the strong field
seeking atoms, and they will finally hit the wire.
\bibitem{LiMOT}
The atoms are loaded into the MOT for 20 sec out of an effusive
thermal beam at a red laser detuning of 25 MHz and a total beam
power of about 150 mW. An electro-optic modulator produces
sidebands of 812 MHz (30\%) one of which is used as a repumper. To
increase the loading rate by a factor of 5 we use an additional 20
mW slower beam with a red detuning of about 100 MHz directed
through the MOT into the oven.
\bibitem{test}
      In our lab we are currently experimentally testing an atomic beamsplitter that is
      formed by two intersecting current carrying wires.
\end{thebibliography}
\end{document}